# Open Educational Resources from Performance Task using Video Analysis and Modeling - Tracker and K12 science education framework

Loo Kang WEE

Ministry of Education, Educational Technology Division, Singapore

wee_loo_kang@moe.gov.sg

Abstract:
This invited paper discusses why Physics performance task by grade 9 students in Singapore is worth participating in for two reasons; 1) the video analysis and modeling are open access, licensed creative commons attribution for advancing open educational resources in the world and 2) allows students to be like physicists, where the K12 science education framework is adopted.
Personal reflections on how physics education can be made more meaningful in particular Practice 1: Ask Questions, Practice 2: Use Models and Practice 5: Mathematical and Computational Thinking using Video Modeling supported by evidence based data from video analysis.
This paper hopes to spur fellow colleagues to look into open education initiatives such as our Singapore Tracker community open educational resources curate on http://weelookang.blogspot.sg/p/physics-applets-virtual-lab.html as well as digital libraries http://iwant2study.org/lookangejss/ directly accessible through Tracker 4.86, EJSS reader app on Android and iOS and EJS 5.0 authoring toolkit for computer models in Easy Java Simulation-Open Source Physics Project.
Recorded session: http://youtu.be/XgPIp6klPyA
PACS: 01.40.gb 01.50.H– 01.50.ht 01.50.hv 45.50.Dd

## I. Introduction

In today's technologically advanced and internet connected world, free education (Caswell, Henson, Jensen, & Wiley, 2008) is the new 'king', possibly through efforts like massive online open course (Johnson et al., 2012)(MOOC), aimed at unlimited and open participation via the web.

This invited paper is largely the author's personal reflections on an unique effort in Raffles Girls School three hundred Grade 9 students with 3 Physics Teachers, framed using the K12 Science Education Framework (Helen, Heidi, & Thomas, 2012) and the video analysis and modeling tool Tracker (Douglas Brown, 2012a), to encourage students to be more like physicists.

## II. What

Tracker is a video analysis and modeling tool built on the Open Source Physics (OSP) Java framework. Though it is possible to run from the Web start or a 5.6 Mb tracker_8.46.jar file, we recommend using the respective installers found at http://www.cabrillo.edu/~dbrown/Tracker/, especially to enable the Xuggle video engine (Douglas Brown, 2012b) that can decode most video file formats. Installers for Tracker version 4.86 installers are available in Windows, Mac OS X as well as Linux operating systems.

## III. Why

The most key reason why I have used tracker extensively is it has allows me to produce open quality physics educational resources in the form of the video analysis and computer model, to provide open educational resources. Commercial physics educational software, presents heightened barriers to adapt and use at home and schools, such as requiring some payment to purchase and later on possible copyright infringement of commercial products when sharing beyond the terms and conditions of use of these commercial software.

Tracker on the other hand, is a free and open-sourced software and is still continually improved by Douglas Brown and his OSP community, has allow ordinary teachers and students to use the software and distribute the video analysis and models resources for others to learn and benefit from such as the large number of video analysis and models built by the Singapore Tracker community http://iwant2study.org/lookangejss/. The tracker version 4.80 and above files are now packaged as a single compressed *.trz file with all the resources inside, usable and editable with a simple mouse click on the file.

## IV. How

I found the K12 Science Education Framework (Figure 1) to be a comprehensive tool to promote and guide students as scientists and the paper reflects on how these 3 practices below (IV.A, IV.B, and IV.C) can be used to further deepen becoming like physicists.

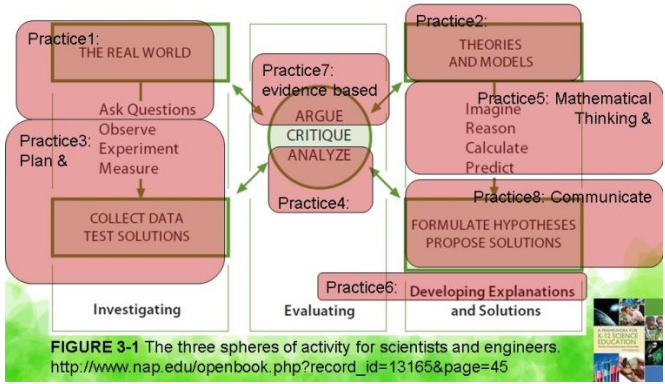

Figure 1. K12 Science Education Framework use to situate the use of Tracker and becoming more like scientists.

### A. *Practice 1: Ask Question situated in the real world that are Mode-able or Complex*

Getting grade 9 students to ask and research on their own physics questions is a difficult task thus a possible scaffold could be guiding them to ask questions that are model-able (Figure 2) using the tracker's dynamics (Loo Kang Wee, Chew, Goh, Tan, & Lee, 2012) or kinematics (Loo Kang Wee, 2012) model.

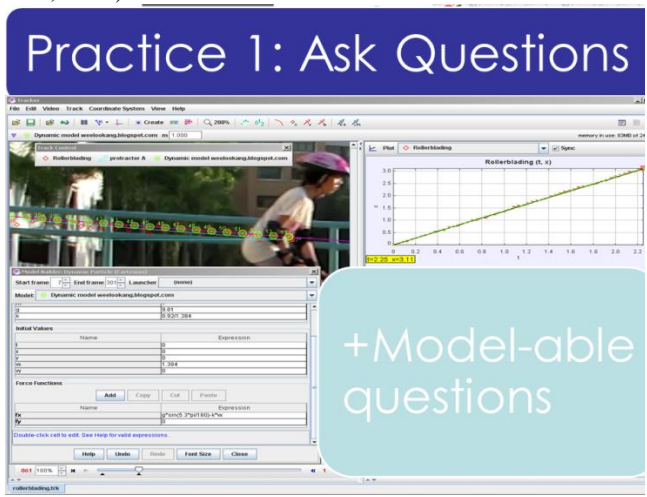

Figure 2. Practice 1: Ask Questions that are model-able in tracker as a possible scaffold to guide students

Complex questions suited for Tracker use could also be like a moving reference frame, for example, a video footage filmed on a moving boat on 2 other sailing boats (Figure 3), while using the constant velocity boat as a moving reference to determine the kinematics graph of the boat of interest.

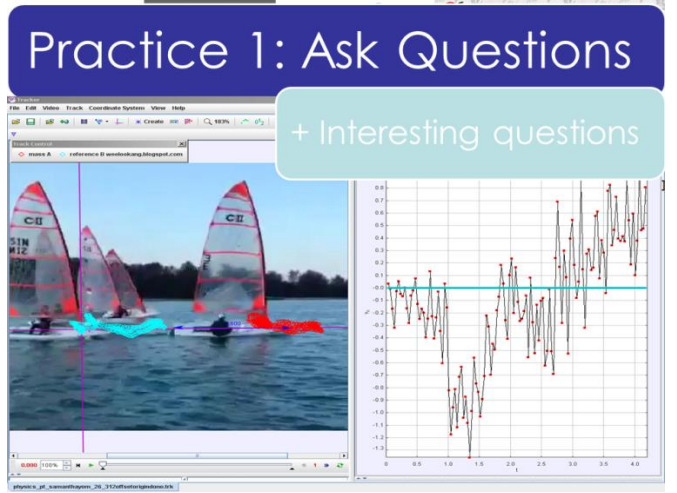

Figure 3. Practice 1: Ask Questions that are complex in tracker as a possible interesting inquiry to motivate students. Notice the velocity of the boat on the right is calculated based on the difference between the boat of the left (moving reference frame).

### B. *Practice 2: Use Models, Easy Java Simulations(EJS) models?*

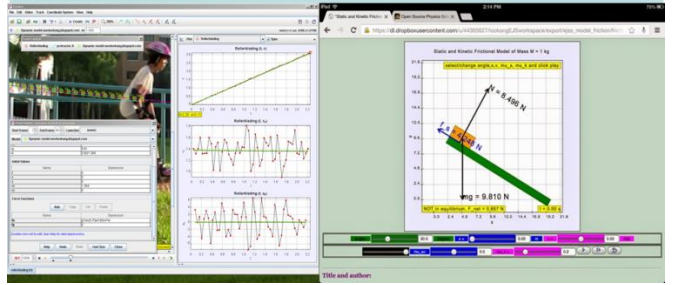

Figure 4. Practice 2: Use Models such as existing Easy Java Simulation models for example the sliding frictional model on the right is available https://dl.dropboxusercontent.com/u/44365627/lookangEJSworkspace/export/ejss_model_friction/friction_Simulation.html

Though this particular understanding of practice 2 is not well enacted, in the form of using existing EJS computer models (Christian & Esquembre, 2012; L.K. Wee & Lye, 2012; Loo Kang Wee, Goh, & Lim, 2014; L. K. L. Wee et al., 2012), because of the difficulties of finding models (Figure 4) that can be used, one possible way to be more like scientists as real scientists and engineers use simulations routinely to test their questions or hypothesis.

Another possible way to strengthen practice involves Tracker video modeling where student begin with initial model of how the motion is govern by to later on progress and improve the model to include and account for damping or frictional forces the evidences suggests that it is applicable.

Overseas Chinese Physicists and Astronomers 23-27 June 2014 OCPA8 Invited Paper, Nanyang Technological University, Singapore

## C. Practice 5: Mathematical and Computational Thinking, Tracker Video Modeling

The real video recorded by the student is the motion of a tennis ball falling down under gravity (filmed using a hand phone in portrait mode resulting in a video that the falling $y$ direction is horizontal), impact the water and subsequent motion with drag and up thrust and eventually the ball bounce out of the water surface. A kinematics model can be used to promote Practice 5 with both mathematical thinking through video analysis and computational thinking through the model building process elaborated below.

### 1) Free Fall Model

#### a) Mathematical Thinking

The free fall of a tennis ball can be mathematically determined through traditional video analysis of the $y$ vs $t$ graph. In the DataTool, select the region of data that corresponds to the free fall, from $t = 0.000$ to $0.267$ s and a Parabola Fit allows students to determine a Fit Equation of y = $A*t\^2+B*t+C$ where parameter $A$ = -2.317, $B$ = -0.1976 and $C$ = 0.2310 (Figure 5). In Kinematics terms, $A = 0.5\ g$ implying gravitational acceleration $g$ = -4.634 $m/s^2$ (probably due to a video frame rate incorrectly understood as the student could have slow the video down to 30 frame per second instead of the true frame rate), initial velocity $u = B$ = -0.1976 $m/s$ and initial height $y0 = C$ = 0.2310 $m$.

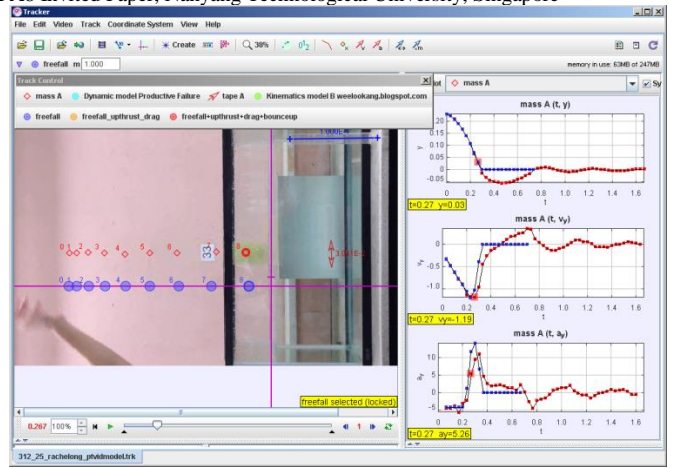

Figure 6. Tracker view of the real data (RED) and the free fall model (BLUE) from $t = 0.000$ to $0.267$ $s$ allows students to see how the model they have proposed matches the real world data and the 3 scientific graphs of $y$ vs $t$, $vy$ vs $t$ and $ay$ vs $t$ (RIGHT).

### 2) Free Fall, Upthrust and Drag Model

#### a) Mathematical Thinking

After the student is convinced that the real motion can be represented by the free fall model proposed, the student can be guided to look at the motion under the water surface.

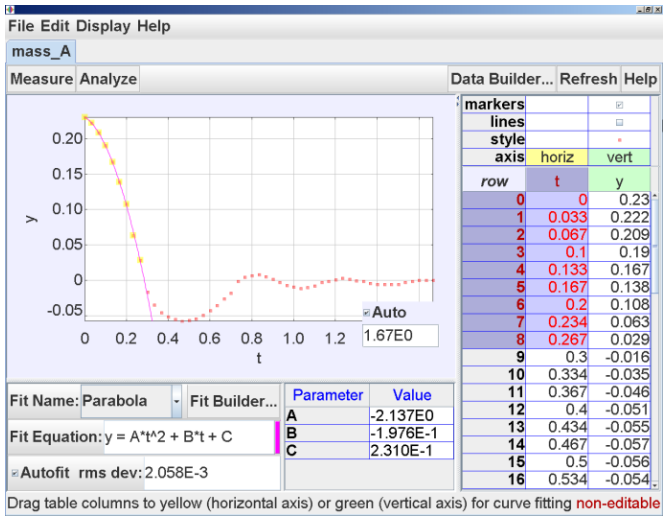

Figure 5. DataTool in Tracker showing from $t = 0.000$ to $0.267$ $s$ the Parabola Fit allows students to determine a Fit Equation of y = $A*t\^2+B*t+C$ where parameter $A$ = -2.317, $B$ = -0.1976 and $C$ = 0.2310.

#### b) Computational Thinking

The mathematical values of the analysis can be used to build the kinematics model by inserting the computational line.

y = if ( t<0.29,(-2.137E0)*t^2-1.976E-1*t+2.310E-1,0)

It means if time $t$ is less than 0.29 $s$, y = (–2.137E0)*t^2– 1.976E-1*t+2.310E-1, else $y$ = 0 , which results in a model view shown in Figure 6.

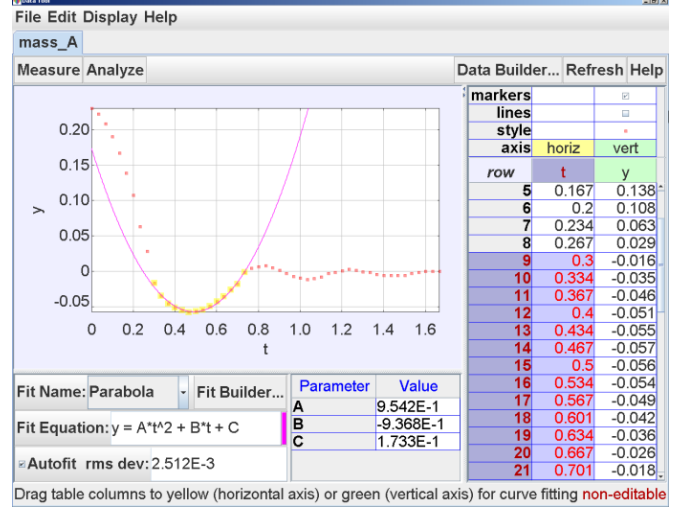

Figure 7. DataTool in Tracker showing from $t = 0.300$ to $0.734$ $s$ the Parabola Fit allows students to determine a Fit Equation of y = $A*t\^2+B*t+C$ where parameter $A$ = –0.9542, $B$ = –0.9368 and $C$ = 0.1733.

Similarly, the student will need to select $t$ = 0.300 to 0.734 s and a Parabola Fit to determine a Fit Equation of y = $A*t\^2+B*t+C$ where parameter $A$ = –0.9542, $B$ = –0.9368 and $C$ = 0.1733 (Figure 7).

In physics terms, the nett force $F_{net}$ is $m*2A$ and equal to the weight $mg$, minus the upthrust $F_{Upthrust}$ and minus the drag force, $F_{Drag}$ as in Equation (1).

$$F_{net} = m(2A) = mg - F_{Upthrust} - F_{Drag} \quad (1)$$

#### b) Computational Thinking

The mathematical values of this analysis can be used to build the kinematics model by inserting the computational line.

3/5

if(t<0.29,(-2.137E0)*t^2-1.976E-1*t+2.310E-1,if(t<0.735,9.542E-1*(t)^2-9.368E-1*(t)+0.17,0))

This is a nested if statement for the meaning of if time *t* is less than 0.29 s, y = (-2.137E0)*t^2-1.976E-1*t+2.310E-1, else if *t* is less than 0.735 s, y = 0.735,9.542E-1*(*t*)^2-9.368E-1*(*t*)+0.17, else y = 0 which results in a model view shown in Figure 8.

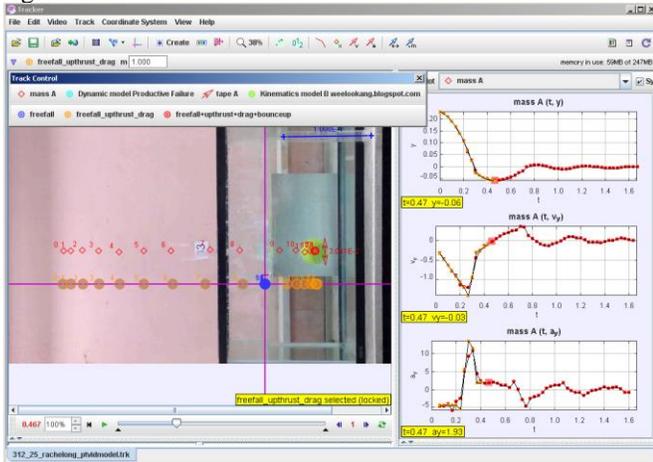

Figure 8. Tracker view of the real data (RED) and the free fall model, upthrust and drag model (ORANGE) from *t* = 0.000 to 0.734 *s* ,allows students to see how the model they have proposed matches the real world data and the 3 scientific graphs of *y* vs *t*, *vy* vs *t* and *ay* vs *t* (RIGHT).

### 3) Free Fall, Upthrust and Drag and Bounce up Model

#### a) Mathematical Thinking

Similarly to part 1) from *t* = 0.67 to 0.901 *s* and a Parabola Fit allows students to determine a Fit Equation of y = A*t^2+B*t+C where parameter A = -0.9661, B = 1.592 and C =-0.6479 (Figure 9).

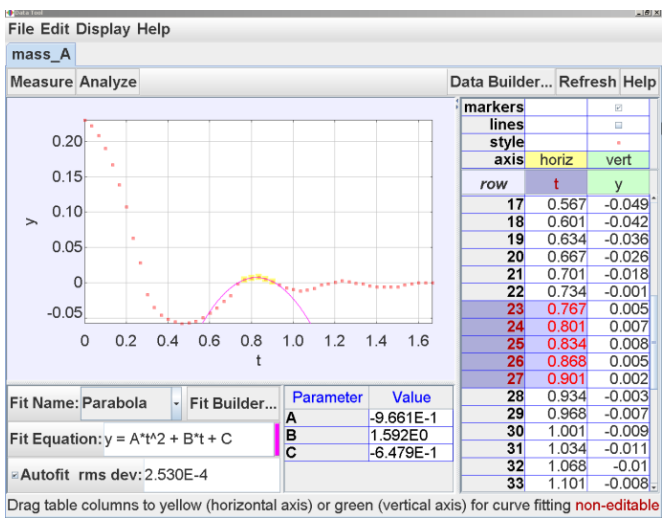

Figure 9. DataTool in Tracker showing from *t* = 0.67 to 0.901 *s* the Parabola Fit allows students to determine a Fit Equation of y = $A*t^2+B*t+C$ where parameter A = -0.9661, B = 1.592 and C =-0.6479.

#### b) Computational Thinking

The mathematical values of the analysis can be used to build the kinematics model by inserting the computational line

if(t<0.29,(-2.137E0)*t^2-1.976E-1*t+2.310E-1,if(t<0.735,9.542E-1*(t)^2-9.368E-1*(t)+0.17,if(t<0.902,-9.661E-1*t^2+1.592E0*t-6.479E-1,0)))

which is a nested if statement for the meaning of if time *t* is less than 0.29 s, y = (-2.137E0)*t^2-1.976E-1*t+2.310E-1, else if *t* is less than 0.735 s, y = 0.735,9.542E-1*(*t*)^2-9.368E-1*(*t*)+0.17, else if *t* is less than 0.902 s, y = 9.661E-1*t^2+1.592E0*t-6.479E-1, else y = 0 which results in a model view shown in Figure 10.

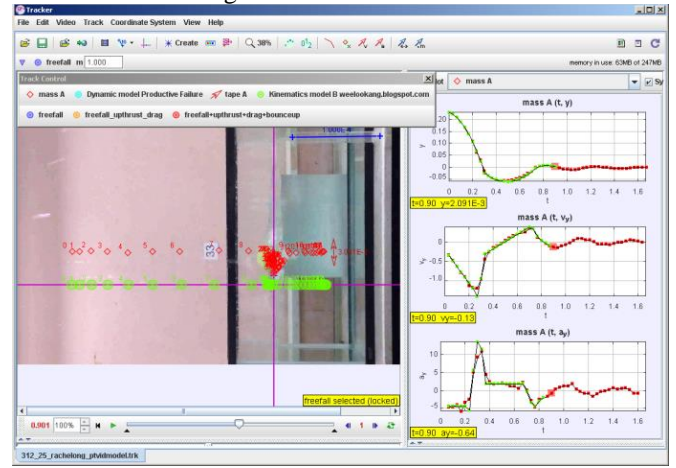

Figure 10. Tracker view of the real data (RED) and the free fall+upthrust+drag+bounceup model (Green) from *t* = 0.000 to 0.901 *s* allows students to see how the model they have proposed matches the real world data and the 3 scientific graphs of *y* vs *t*, *vy* vs *t* and *ay* vs *t* .

Clearly, Tracker's video modeling can afford for both mathematical and computational thinking integrated seamlessly and allow students to eventually progress to even propose a Dynamics Model which would involve terms in equation (1), outside the scope of this paper.

## V. CONCLUSION

The What – Tracker (Douglas Brown, 2012a), Why – Open Educational Resources (2008) and How – K12 Science Education Framework (Helen et al., 2012) are briefly discussed in the context of using video analysis (Loo Kang Wee & Lee, 2011) to allow students to be more like physicists.

Some ideas on how Practice 1 Ask Question, Practice 2 Use Models and Practice 5, Mathematical and Computational thinking.

The sophisticated but meaningful video modeling pedagogy (Doug Brown & Christian, 2011; Douglas Brown, 2009), critical to the fifth practice of science (physics) especially when driven with data from video analysis, allow students to discover using evidences and incomplete models proposed by themselves, to incrementally and iteratively improve and self-invent a better model to predict and explain the motion of their chosen question.

This paper hopes to spur fellow colleagues to look into open education initiatives such as our Singapore Tracker community open educational resources curate on http://weelookang.blogspot.sg/p/physics-applets-virtual-

lab.html as well as digital libraries http://iwant2study.org/lookangejss/ directly accessible through Tracker 4.86, EJSS reader app on Android and iOS and EJS 5.0 authoring toolkit for computer models in Easy Java Simulation-Open Source Physics Project.

ACKNOWLEDGMENT

We wish to acknowledge the passionate contributions of Douglas Brown, Wolfgang Christian, Mario Belloni, Anne Cox, Francisco Esquembre, Harvey Gould, Bill Junkin, Aaron Titus and Jan Tobochnik for their creation of Tracker video analysis and modeling tool.

The insights for this paper is made possible; thanks to the eduLab project NRF2013-EDU001-EL07 Becoming like Scientists through video analysis, awarded by the National Research Foundation, Singapore in collaboration with National Institute of Education, Singapore and the Ministry of Education (MOE), Singapore.

Any opinions, findings, conclusions or recommendations expressed in this paper, are those of the author and do not necessarily reflect the views of the MOE, NIE or NRF.

REFERENCE

Brown, Doug, & Christian, Wolfgang. (2011, Sept 15-17). *Simulating What You See.* Paper presented at the MPTL 16 and HSCI 2011, Ljubljana, Slovenia.

Brown, Douglas. (2009). *Video Modeling with Tracker*. Paper presented at the American Association of Physics Teachers AAPT Summer Meeting, Ann Arbor. http://cabrillo.edu/~dbrown/tracker/video_modeling.pdf

Brown, Douglas. (2012a). Tracker Free Video Analysis and Modeling Tool for Physics Education. from http://www.cabrillo.edu/~dbrown/tracker/

Brown, Douglas. (2012b). Xuggle Installers http://www.cabrillo.edu/~dbrown/tracker/. from http://www.cabrillo.edu/~dbrown/tracker/

Caswell, Tom, Henson, Shelley, Jensen, Marion, & Wiley, David. (2008). Open content and open educational resources: Enabling universal education. *The International Review of Research in Open and Distance Learning, 9*(1), 11.

Christian, Wolfgang, & Esquembre, Francisco. (2012, Jul 04, 2011 - Jul 06, 2011). *Computational Modeling with Open Source Physics and Easy Java Simulations.* Paper presented at the South African National Institute for Theoretical Physics Event, University of Pretoria, South Africa.

Helen, Quinn, Heidi, Schweingruber, & Thomas, Keller. (2012). *A Framework for K-12 Science Education: Practices, Crosscutting Concepts, and Core Ideas*. Washington, DC: The National Academies Press.

Johnson, L., Adams, S., Becker, S., Ludgate, H., Cummins, M., & Estrada, V. (2012). Technology Outlook for Singaporean K-12 Education 2012-2017: An NMC Horizon Project Regional Analysis. Austin, Texas: New Media Consortium.

Resources, Open Educational. (2008). Open Educational Resources. Retrieved 02 June, 2008, from http://www.oercommons.org/

Wee, L.K., & Lye, S.Y. (2012). *Designing Open Source Computer Models for Physics by Inquiry using Easy Java Simulation*. Paper presented at the 20th International Conference on Computers in Education (ICCE 2012) Singapore Interactive Event, Singapore. http://arxiv.org/ftp/arxiv/papers/1210/1210.3412.pdf

Wee, Loo Kang. (2012). Tracker Video Analysis: Bouncing Ball. Retrieved from http://www.compadre.org/Repository/document/ServeFile.cfm?ID=11705&DocID=2583

Wee, Loo Kang, Chew, Charles, Goh, Giam Hwee, Tan, Samuel, & Lee, Tat Leong. (2012). Using Tracker as a pedagogical tool for understanding projectile motion. *Physics Education, 47*(4), 448.

Wee, Loo Kang, Goh, Giam Hwee, & Lim, Ee-Peow. (2014). *Easy Java Simulation, an innovative tool for teacher as designers of gravity-physics computer models*. Paper presented at the Multimedia Physics Teaching and Learning Conference Madrid, Spain. http://arxiv.org/ftp/arxiv/papers/1401/1401.3061.pdf

Wee, Loo Kang Lawrence, Lim, Ai Phing, Goh, Khoon Song Aloysius, Lye, Sze Yee, Lee, Tat Leong, Xu, Weiming, . . . Lim, Kenneth Y T. (2012). *Computer Models Design for Teaching and Learning using Easy Java Simulation* Paper presented at the The World Conference on Physics Education Istanbul, Turkey.

Wee, Loo Kang, & Lee, Tat Leong. (2011). *Video Analysis and Modeling Tool for Physics Education: A workshop.* Paper presented at the 4th Redesigning Pedagogy conference, National Institute of Education, Nanyang Technological University, Singapore. http://conference.nie.edu.sg/2011/papers_pdf/WOR074.pdf

AUTHOR

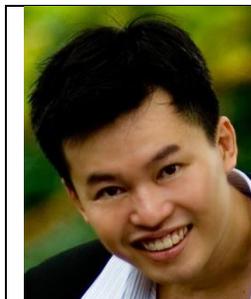

Loo Kang WEE Lawrence is currently educational technology specialist II at the Ministry of Education, Singapore. He was a junior college physics lecturer and his research interest is in Open Source Physics tools like Easy Java Simulation for designing computer models and use of Tracker. His contributions garner awards including Public Service PS21 Distinguished Star Service Award 2014 and Best Ideator 2012, Ministry of Education, Best Innovator Award 2013 and Excellence Service Award 2012.